\documentclass[journal=jctcce,manuscript=article]{achemso}

\usepackage{mciteplus}
\usepackage{graphicx}
\usepackage{amsmath}
\usepackage{amsfonts}
\usepackage{booktabs}
\usepackage{float}

\renewcommand{\r}{\boldsymbol{r}}
\newcommand{\q}{\boldsymbol{q}}
\renewcommand{\l}{\boldsymbol{l}}
\newcommand{\z}{\boldsymbol{z}}
\newcommand{\p}{\boldsymbol{p}}
\newcommand{\pq}{\boldsymbol{p}_q}
\newcommand{\pc}{p_{\mu}}
\newcommand{\pl}{\boldsymbol{p}_{l}}
\newcommand{\I}{\boldsymbol{I}}
\newcommand{\M}{\boldsymbol{M}}
\newcommand{\Ml}{\boldsymbol{M}_{l}}
\newcommand{\Mq}{\boldsymbol{M}_{q}}
\newcommand{\mq}{m_q}
\newcommand{\mc}{m_{\mu}}
\newcommand{\Gl}{\boldsymbol{\Gamma}}
\newcommand{\gq}{\gamma_q}
\newcommand{\gc}{\gamma_{\mu}}
\newcommand{\Tl}{\boldsymbol{T}}
\newcommand{\Tq}{T_q}
\newcommand{\Tc}{T_{\mu}}
\newcommand{\s}{\boldsymbol{s}}
\renewcommand{\t}{\boldsymbol{t}}

\newcommand{\J}{\boldsymbol{J}}
\newcommand{\h}{\boldsymbol{\chi}}
\newcommand{\0}{\boldsymbol{0}}
\newcommand{\1}{\boldsymbol{1}}
\newcommand{\A}{\boldsymbol{A}}
\renewcommand{\b}{\boldsymbol{b}}

\newcommand{\f}[2]{\frac{\partial #1}{\partial #2}}
\newcommand{\bm}[4]{\begin{pmatrix}#1&#2\\#3&#4\end{pmatrix}}
\newcommand{\bv}[2]{\begin{pmatrix}#1\\#2\end{pmatrix}}

\newcommand{\ene}{~\mathrm{kcal~mol^{-1}}}
\newcommand{\enea}{~\mathrm{kcal~mol^{-1}~atom^{-1}}}
\newcommand{\tim}{~\mathrm{fs}}
\newcommand{\tima}{~\mathrm{ps}}
\newcommand{\timb}{~\mathrm{ns}}
\newcommand{\tem}{~\mathrm K}

\title{\Large{Stochastic Constrained Extended System Dynamics for Solving Charge Equilibration Models}}

\author{\large{Songchen Tan$^{1,2}$, Itai Leven$^{2,3}$,Dong An$^{4,5}$, Lin Lin$^{4,5}$, Teresa Head-Gordon$^{2,3,6}$}}
\affiliation{$^1$College of Chemistry and Molecular Engineering, Peking University, China}
\affiliation{$^2$Kenneth S. Pitzer Theory Center,  University of California, Berkeley, CA, USA}
\affiliation{$^3$Chemical Sciences, Lawrence Berkeley National Laboratory, Berkeley, CA, USA}
\affiliation{$^4$Department of Mathematics, University of California, Berkeley, CA, USA}
\affiliation{$^5$Computational Research, Lawrence Berkeley National Laboratory, Berkeley, CA, USA}
\affiliation{$^6$Departments of Chemistry,  Bioengineering, and Chemical and Biomolecular Engineering, University of California, Berkeley, CA, USA, }
\email{thg@berkeley.edu}
\begin{document}
\maketitle
\begin{abstract}
\noindent
We present a new stochastic extended Lagrangian solution to charge equilibration that eliminates self-consistent field (SCF) calculations, eliminating the computational bottleneck in solving the many-body solution with standard SCF solvers. By formulating both charges and chemical potential as latent variables, and introducing a holonomic constraint that satisfies charge conservation, the SC-XLMD method accurately reproduces structural, thermodynamic, and dynamics properties using ReaxFF, and shows excellent weak- and strong-scaling performance in the LAMMPS molecular simulation package.
\end{abstract}

\section{Introduction}
Many non-reactive and reactive force fields have relied on the electronegativity equalization method\cite{Mortier86,Mortier85} (EEM) or charge equilibration method\cite{Rappe91} (CEM) to describe charge flow in and between molecules. The EEM approach was inspired by the concepts of atomic electronegativity and hardness drawn from Density Functional Theory\cite{Parr83} to define an electrostatic model that allows the charges on atoms to fluctuate with changing nuclear configurations during molecular simulations.\cite{Mortier86,Mortier85} The EEM approach was later generalized to CEM by Rappe and Goddard to include a screened electrostatic interaction between charges, using empirical parameters such as atomic ionization potentials, electron affinities, and atomic radii to parameterize the model.\cite{Rappe91} The CEM model has been successfully applied to a variety of chemical systems such as proteins\cite{Patel04} and membranes\cite{Bauer12}, metal-organic frameworks,\cite{Wilmer11,Wilmer13} and to describe the quartz-stishovite phase transition.\cite{Demiralp99} 

The rate limiting step of the CEM (and EEM) model is the determination of the new charges from two sets of linear equations, which represent the minimization of the total energy for the new nuclear configuration under a constraint that the total charge of the system is conserved. This may be solved directly for small systems (typically with Cholesky decomposition), but must be solved iteratively in practice for large systems using solvers such as the direct inversion in the iterative subspace (DIIS)\cite{Pulay80} or conjugate gradient (CG) methods.\cite{Wang05} The number of self-consistent field (SCF) iterations can be reduced with careful preconditioning, polynomial extrapolation from previous steps, and good software implementations\cite{Nakano1997,Aktulga12,Aktulga19}, but the solution of the many-body CEM forces at each time step remains the most computationally demanding component of MD simulations using ReaxFF\cite{vanduin01}, which is approximately one to two orders of magnitude slower than traditional non-reactive force fields. 

An alternative approach is to formulate an extended system of auxiliary electronic variables that are evolved in time with extended Lagrangian molecular dynamics (XLMD).\cite{Niklasson06,Albaugh15,Albaugh17,Leven19} With an extended Lagrangian that includes fictitious kinetic energy of auxiliary charges, as well as a potential energy that keep the auxiliary charges close to the exact solution, the extended system charges are evolved dynamically using symplectic and time-reversible algorithms to replace the iterative solution for many-body forces. Relevant to CEM, Leven and Head-Gordon have used the dynamically evolved auxiliary charges as an initial guess for the CG-SCF procedure, thereby allowing for a more loose convergence tolerance for final charges without introducing additional (and sometimes even diminishing) energy drift which measures the stability of the simulation.\cite{Leven19} The resulting inertial extended Lagrangian SCF (iEL/SCF) method was shown to successfully reduce the number of SCF iterations by half or more in the CEM solutions for the reactive force field ReaxFF\cite{vanduin01,Leven19}. 

In this work we further extend the iEL/SCF method for CEM by eliminating SCF cycles altogether, as we have done previously for non-reactive force field models using iEL/0-SCF.\cite{Albaugh17,An2020} For polarizable models, the auxiliary induced dipoles evolve under a harmonic potential that keeps their values close to the converged real dipole solution, as approximated by a one-time step estimation derived from a local-kernel mixing of the real and auxiliary variables using an optimal mixing parameter $\gamma$. This SCF-free approximation works well if the real dipole dynamics evolve on a longer timescale, well-separated from the discretized time step, so that a local-kernel mixing remains a good approximation to the true SCF solution. A second important consideration is to control the problem of resonances, i.e. errors in the time-integration of the harmonic forces that leak to the auxiliary kinetic energy and create numerical instabilities, but which can be controlled through a separate thermostat for the auxiliary variables as we have shown previously.\cite{Albaugh17,An2020} Recently, An and co-workers have developed a new formulation of an iteration-free scheme, Stochastic-XLMD, where a thermostat coupling parameter $\varepsilon$ replaces the mixing parameter $\gamma$, and the effect of a Langevin thermostat applied to the latent induced dipole variables for classical polarization was shown to be robust, although not strictly tiem-reversible.\cite{An2020}

However, the generalization of the resulting iEL/0-SCF method from induced dipoles to fluctuating charges is not straightforward for three reasons: (1) the characteristic decorrelation time for charges is more than one order of magnitude faster than for induced dipoles, (2) the charges are derived under a constraint that the net charge of the entire system is conserved, and (3) the resonance problem may be more severe under a harmonic potential now applied to two sets of coupled linear equations. To illustrate, an iteration-free XLMD scheme for CEM\cite{Lipparini11, Nomura15}, was found to be unstable after time propagation of no more than several picoseconds,\cite{Leven19} which is generally not sufficient for converging thermodynamic quantities. 

In this work, we have addressed these issues through careful formulation of an XLMD procedure that utilizes two latent variables - the charge and chemical potential - and enforces the conservation of charge through a holonomic constraint scheme that is conforming for both energy and forces\cite{Zhang19}. We have combined this new SCF-less solution for CEM with the Stochastic-XLMD (SXLMD) method for thermostatting\cite{An2020}, and implemented it within the ReaxFF force field in the Large-scale Atomic and Molecular Massive Parallel Simulation (LAMMPS) library\cite{Plimpton95}. We show that the stochastic and constrained extended Lagrangian scheme with no iteration, SC-XLMD, is capable of producing stable trajectories over a timescale of nanoseconds, while retaining energy conservation and equivalent thermodynamics properties of the typical CG-SCF solution. Compared to the standard implementation in LAMMPS, the computational speed of the new SC-XLMD approach is comparable to a fixed charge calculation, and scales linearly with increasing number of cores or increasing size of simulated systems. 
\section{Theory}
\textit{Charge Equilibration Equation.}
A general form of the Hamiltonian for the CEM (and EEM) model is
\begin{equation}
\begin{aligned}
H&=\frac12\p^T\M^{-1}\p+U(\r)+V(\r,\q)
\end{aligned}
\end{equation}
where $\r\in\mathbb R^{3n}$ and $\p\in\mathbb R^{3n}$ are the atom positions and momenta, $\M=\operatorname{diag}\{m_1\I_3,\cdots,m_n\I_3\}$ are the atom mass (diagonal) matrix, $U(\r)$ encompasses molecular interactions other than the many-body electrostatic potential, $V(\r,\q)$,
\begin{equation}
\begin{aligned}
V(\r,\q)=\frac12\q^T\J(\r)\q+\h^T\q
\end{aligned}
\end{equation}
which is the focus of this work. The potential term in Eq. (2) describes changes in the charges $\q$ that are dependent on the electronegativity of atoms when bearing zero charge, $\h$, and the screened electrostatic interaction matrix, $\J$, comprised of the following matrix elements 
\begin{equation}
J_{ij}=\delta_{ij}\eta_i+(1-\delta_{ij})(r_{ij}^3+\gamma_{ij}^{-3})^{-1/3}
\end{equation}
where $\eta_i$ is related to the atomic hardness, $\gamma_{ij}$ is the electrostatic screening parameter, and $r_{ij}$ is the distance between atoms $i$ and $j$.

The CEM model allows the charge to rearrange according to the minimization of the potential energy $V(\r,\q)$ as a response to the motion of atoms, while enforcing the constraint that the sum of charges remains constant (without loss of generality, we assume the sum of charges is 0 henceforth):
\begin{equation}
\begin{split}
L(\r,\q,\mu)&=V(\r,\q)-\mu\1^T\q\\
&=\frac12\q^T\J(\r)\q+\h^T\q-\mu\1^T\q
\end{split}
\end{equation}
and
\begin{equation}
\f{L}{\q}=0\quad \f{L}{\mu}=0
\end{equation}
where the Lagrange multiplier $\mu$ is the chemical potential. This yields an $n+1$-dimensional equation:
\begin{equation}
\bm{\J}{-\1}{-\1}{0}\bv{\q}{\mu}=\bv{-\h}{0}
\end{equation}
Since a direct inversion of this matrix is prohibitive in practice, Eq. (6) is instead solved by partitioning the original charges $\q$ into fictitious charges $\s$ and $\t$, $\q=\s-\mu\t$, and then solving $\J\s=-\h$ and $\J\t=-\1$ iteratively with a conjugate gradient (CG) method.\cite{Aktulga12} The number of iterations can be reduced with careful preconditioning, as well as polynomial extrapolation from previous steps, but the overall computational cost is still significantly larger than conventional simulations using fixed charges, and defines the rate limiting step for reactive force field simulations of large systems.

\textit{Extended System Dynamics: the SC-XLMD Method.} An alternative solution to the charge equilibration problem is to formulate an extended system by introducing latent variables that evolve in time with the real degrees of freedom using an XLMD algorithm, as shown in many previous studies.\cite{Niklasson06,Albaugh15,Albaugh17}. However, to the best of our knowledge, there is no known example of successfully treating the charge conservation constraint and the time-evolution consistently for the case of fluctuating charges. 

One class of XLMD method\cite{Nomura15,Li19} is to assign latent momenta $\pq$ and latent mass $\Mq=\mq \I_n$ to the corresponding $\q$ variables, and utilizing a Hamiltonian of the form

\begin{equation}
\begin{aligned}
H_{\rm ext}^{(1)}(\r,\p,\q,\pq)&=\frac12\p^T\M^{-1}\p+\frac12\pq^T\Mq^{-1}\pq\\
&+U(\r)+V(\r,\q)
\end{aligned}
\end{equation}
This XLMD approach completely ignores any constraint and thus will exhibit significant problems with charge conservation in a long simulation. 

The second class of XLMD method,\cite{Lipparini11,Rick94} utilizes a Hamiltonian that is equivalent to
\begin{equation}
\begin{aligned}
H_{\rm ext}^{(2)}(\r,\p,\q,\pq)&=\frac12\p^T\M^{-1}\p+\frac12\pq^T\Mq^{-1}\pq\\
&+U(\r)+V(\r,\q)-\mu\1^T\q
\end{aligned}
\end{equation}
where $\mu=\mu(\r,\q)$ is determined on-the-fly by solving the following algebraic equation every time step,
\begin{equation}
    \1^T(\J\q+\h-\mu\1)=0
\end{equation}
However, the resulting differential-algebraic system is non-Hamiltonian and is vulnerable to numerical noise. As a result, both of these classes of XLMD methods have not been able to perform a simulation for CEM longer than $10\tima$, which is generally not enough for converging thermodynamic quantities.

Here instead, we consider a new extended Hamiltonian in which we treat the charges $\q$ and chemical potential $\mu$ together as an extended set of latent positions, $\l=(\q,\mu)$ with latent momenta $\pl=(\pq,\pc)$ as well as latent mass $\Ml=\operatorname{diag}\{\mq \I_n,\mc\}$
\begin{equation}
\begin{aligned}
H_{\rm ext}^{(3)}(\r,\p,\l,\pl)&=\frac12\p^T\M^{-1}\p+\frac12\pl^T\Ml^{-1}\pl\\
&+U(\r)+\frac12\l^T\A(\r)\l-\b^T\l
\end{aligned}
\end{equation}
where 
\begin{equation}
\A(\r)=\bm{\J(\r)}{-\1}{-\1}{0}\quad\b=\bv{-\h}{0}
\end{equation}
Since the many-body potential term in $H_{\rm ext}^{(3)}$ is exactly what is minimized in the Lagrange multiplier method, the evolution of the latent variables will consistently keep close to the Born-Oppenheimer energy surface, as well as keeping the total charge constant. Furthermore, by making connections to the well-known holonomic constraint scheme in classical molecular dynamics,\cite{Zhang19} we can define a function $\z(\l)$ of the latent variables described as a projection,
\begin{equation}
\z(\l) = \bm{\I_n-\1\1^T/n}{0}{0}{1}\l
\end{equation}

By replacing $\l$ in $H_{\rm ext}^{(3)}$ with $\z(\l)$, we arrive at our final expression of the extended Hamiltonian used in this work:
\begin{equation}
\begin{aligned}
H_{\rm ext}(\r,\p,\l,\pl)&=\frac12\p^T\M^{-1}\p+\frac12\pl^T\Ml^{-1}\pl\\
&+U(\r)+\frac12\z^T(\l)\A(\r)\z(\l)-\b^T\z(\l)
\end{aligned}
\end{equation}
We note that when we derive the latent force from $H_{\rm ext}$, we obtain
\begin{equation}
\dot{\pl} = -\f{H^*_{\rm ext}}{\l} = \bm{\I_n-\1\1^T/n}{0}{0}{1} (\b-\A\z)
\end{equation}
which satisfies $\1^T\dot\pq=0$. Therefore, with proper initialization of the latent position and momenta that satisfies $\1^T\q(0)=\1^T\pq(0)=0$, the time evolution keeps $\1^T\q(t)=0$ for arbitrary $t$. We note that implementing this constraint strategy has been successful for retaining the symplectic structure of integration, which is beneficial for long time stable simulation.\cite{Zhang19}

\section{Methods}
\textit{Langevin Thermostat and Integration Algorithm}. We utilize the Langevin thermostat approach we have developed previously for thermostating polarizable models\cite{An2020}, which requires us to define both dissipation $\Gl=\operatorname{diag}\{\gq \I_n,\gc\}$ and temperature $\Tl=\operatorname{diag}\{\Tq \I_n,\Tc\}$ parameters corresponding to the latent variables. Once defined, the "BAOAB" scheme is used to achieve efficient thermostatting\cite{Li17} by propagating the equations of the extended system as follows:

\begin{itemize}
    \item Step B for $\Delta t/2$: 
    $$
    \p\leftarrow\p-\f{H^*}{\r}\frac{\Delta t}2
    $$
    $$
    \pl\leftarrow\pl-\f{H^*}{\l}\frac{\Delta t}2
    $$
    \item Step A for $\Delta t/2$:
    $$
    \r\leftarrow\r+\M^{-1}\p\frac{\Delta t}2
    $$
    $$
    \l\leftarrow\l+\Ml^{-1}\pl\frac{\Delta t}2
    $$
    \item Step O for $\Delta t$:
    $$
    \pl\leftarrow e^{\Gl\Delta t}\pl+\sqrt{1-e^{2\Gl\Delta t}}\sqrt{\Ml\Tl}\boldsymbol{\xi}
    $$
    where $\boldsymbol{\xi}$ is a vector consists of $n+1$-dimensional independent standard normal random variable.
    \item Step A for $\Delta t/2$;
    \item Step B for $\Delta t/2$; 
\end{itemize}
Our previous work proved that the trajectory of the extended system will converge to the real system when $\Ml\to0$ for arbitrary initial condition of latent variable $\l$.\cite{An2020} 

Having formulated the integration algorithm with thermostats, we now explain the rationale to determine the parameters. We first note that for a one-dimensional harmonic oscillator with force constant $k$ and mass $m$, a stable numerical integration should satisfy $\Delta t^2k/m=\Delta t^2\omega^2<2$.\cite{Niklasson2008} For the CEM model, the ``force constant'' is $\J(\r)$, thus it is subject to
\begin{equation}
\Delta t^2\rho(\J(\r))\mq^{-1}<2
\end{equation}
where $\rho(\cdot)$ denotes the 2-norm of a matrix. For the symmetric matrix $\J(\r)$, this is the maximum of the absolute of its eigenvalue, which is approximately the inverse of the minimum atom hardness, $\eta_{\min}^{-1}$. The above equation in turn determines the mass by
\begin{equation}
\mq>m_{q,\min}=\frac{\Delta t^2}{2\eta_{\min}}
\end{equation}
In practice, we will use $\mq=5m_{q,\min}$ through the paper to ensure a stable trajectory.


The choice of charge temperature $\Tq$ is determined by taking derivatives on both sides of the charge equilibration equation:
\begin{equation}
\bm{\dot\J}{\0}{\0}{0}\l+\bm{\J}{-\1}{-\1}{0}\dot{\l}=\bv{0}{0}
\end{equation}

where the derivative of $\J$ can be easily calculated by noticing $J_{ij}=f(r_{ij})$, and
\begin{equation}
\frac{\mathrm dr_{ij}}{\mathrm dt}=\sum_{\alpha=x,y,z}\frac{(r_{i\alpha}-r_{j\alpha})(v_{i\alpha}-v_{j\alpha})}{r_{ij}}
\end{equation}
so that
\begin{equation}
\dot J_{ij}=\frac{f'(r_{ij})}{r_{ij}}\sum_{\alpha=x,y,z}(r_{i\alpha}-r_{j\alpha})(v_{i\alpha}-v_{j\alpha})
\end{equation}

In practice, short exact trajectories using a tight convergence ($10^{-12}$) is calculated and the charge temperature $T_q$ is estimated by:
\begin{equation}
T_q=\frac12 \mq\langle\dot{\q}^T\dot{\q}\rangle / n
\end{equation}
In regards the parameters related to chemical potential, i.e. $(\mc, \gc, \Tc)$, we note that in practice the initial velocity $\dot{\mu}$ solved from the above equations were found to be negligible, so we assign $\dot{\mu}(0)=0$, then $\mu$ will be a constant of motion, thus we no longer need to determine them.



\textit{Simulation Methods}. We have implemented SC-XLMD within the framework of the LAMMPS software package\cite{Plimpton95} for ReaxFF\cite{vanduin01}. We use a water box comprising 233 water molecules, for which the force field developed by Rahaman \emph{et al} is used.\cite{Rahaman2010} For all NVE simulations, the system is first equilibrated in the NVT ensemble (with Nosé-Hoover thermostats\cite{Martyna1996}) at $300\tem$ for $10\tima$, followed by NVE propagation for $500\tima$, using a charge mass of $70~\ene\tim^2\mathrm{e^{-2}}$ and temperature of $1.5~\mathrm{K}$ as determined above. For all NVT simulations, in addition to the thermostat applied to the latent variables, the real system variables thermostatted with a 4th-order Nosé-Hoover chain\cite{Martyna1996} at $300\tem$ for $500\tima$. A time step of $\Delta t=0.15\tim$, which is suggested by previous work\cite{Lipparini11} was used in all simulations.

\section{Results}
\textit{Energy Conservation and Latent Temperature.} The conservation of the total energy is the most important intrinsic indicator of correct dynamics for any molecular dynamics algorithm. We note that energy conservation for ReaxFF in LAMMPS may be poor due to discontinuities in the potential energy surface, which has been identified to arise from distance cut-offs of the bond order terms.\cite{Furman2019} Nonetheless we take the standard CG-SCF solution with a $10^{-10}$ convergence criteria as the gold standard for energy conservation comparison. We also consider an XLMD method which uses no latent variable thermostats of the extended system Hamiltonian (C-XLMD), and compare it against the complete SC-XLMD solution which uses Langevin thermostats.

\begin{figure*}[hbt!]
    \centering
    \includegraphics[scale=0.6]{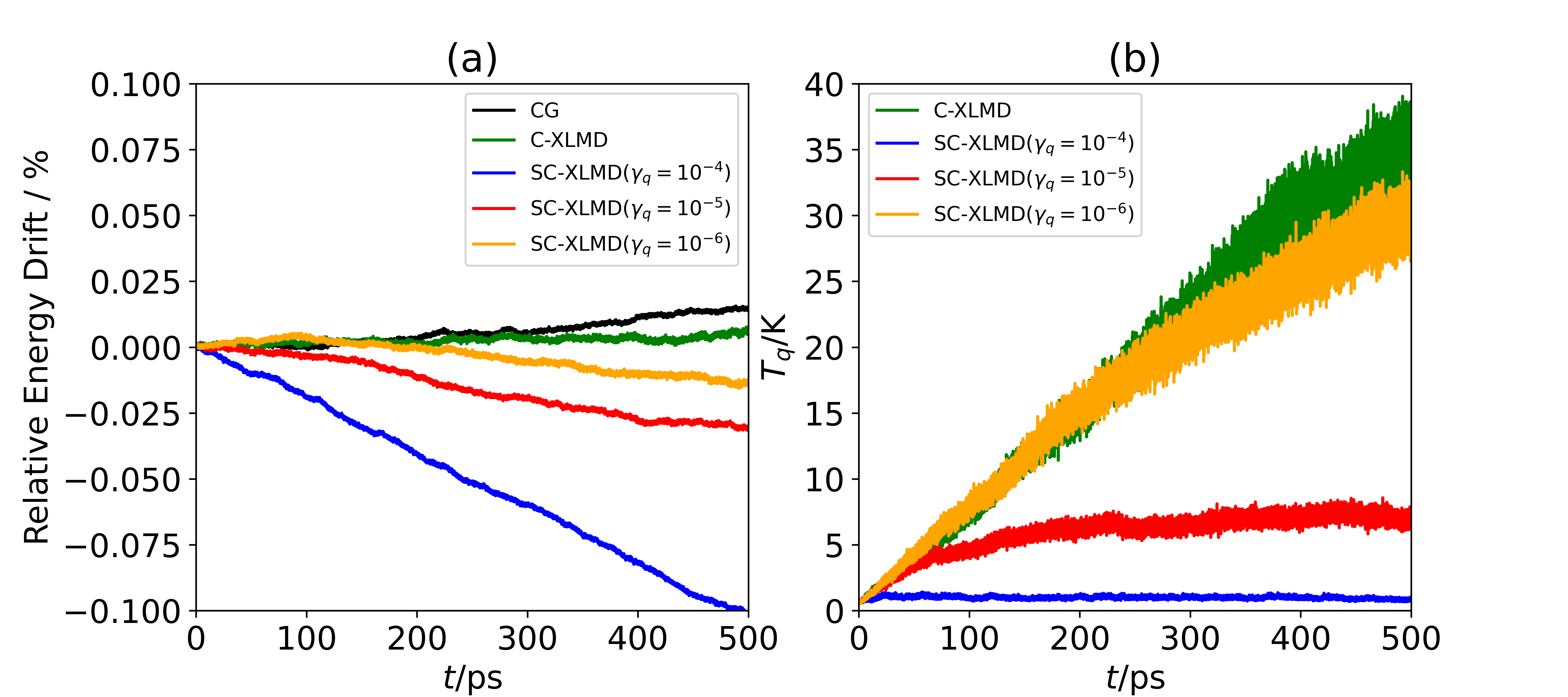}
    \caption{Comparison of methods for energy conservation and charge temperature of the CEM simulation for bulk water. (a) Total relative energy drift in percentage units as a function of time for the standard CG-SCF solution ($10^{-10}$ convergence criteria), an extended Lagrangian with no latent variable thermostating (C-XLMD), and when using SC-XLMD with various values of the latent thermostat coupling parameter $\gamma_q$. The absolute value of energy drift rates in $\mathrm{percent}\timb^{-1}$ are 0.03 (CG-SCF), 0.01 (C-XLMD), 0.22 (SC-XLMD with $\gq=10^{-4}$), 0.05 ($\gq=10^{-5}$) and 0.03 ($\gq=10^{-6}$). (b) The corresponding charge temperature as a function of time for the above methods except CG.}
\end{figure*}

As shown in Figure 1a, there is no more loss of energy conservation for C-XLMD (i.e. with no thermostat coupling $\gq=0$) as compared to the CG-SCF solution. However, as described in previous work, the latent variables are susceptible to numerical noise because of integrator resonance.\cite{Albaugh17,An2020} Without dissipation, the numerical error will accumulate and cause the kinetic energy of latent variables to increase as shown in Figure 1b. The resonance effects will eventually cause degradation of the molecular dynamics that will in turn effect physical observables, hence we require some type of thermostatting of the latent variables at an intrinsically cold temperature to dissipate the error. We thus implemented a Langevin thermostat to control the latent variable temperature as discussed above, but this requires an optimization of the thermostat coupling parameter for SC-XLMD, and we have considered values of $\gq=10^{-4}$, $10^{-5}$ and $10^{-6}$. If $\gq=10^{-4}$, the charge temperature $T_q$ is very stable, but energy conservation is severely compromised with respect to CG-SCF and C-XLMD. On the other hand, if $\gq=10^{-6}$, the energy conservation is very good but the charge temperature $T_q$ will increase over time and properties will be compromised. We therefore suggest that by using $\gq=10^{-5}$, the energy conservation is comparable to the CG-SCF solution and the latent charge temperature is also kept under good control, and is the choice of the thermostat coupling parameter used in the subsequent results.



\textit{Fluctuating Charge Properties.} Next, we assess the ability of C-XLMD and SC-XLMD to produce a similar behavior of charges compared to CG-SCF, which is the key feature of all CEM-based simulations. We examine the charge distribution (a statistic property) as well as the charge autocorrelation function (a dynamic property) defined by
\begin{equation}
\begin{aligned}
C_q(t)&=\frac{\langle (q(t)-\bar q) (q(0)-\bar q)\rangle}{\langle (q(0)-\bar q)^2\rangle}\\
\tilde C_q(\omega)&=\int_0^{\infty}e^{-i\omega t}C_q(t)\mathrm dt
\end{aligned}
\end{equation}

In Figure 2a, while both unthermostated and thermostated XLMD methods are able to qualitatively reproduce the charge distribution generated by CG-SCF, the increase of the charge temperature for C-XLMD has caused a more noticeable dispersion for the charges, while the charge dispersion is better controlled through the Langevin thermostat for SC-XLMD. Figure 2b shows that the increase of the latent charge temperature using the C-XLMD method has given rise to an accelerated decorrelation of real charges compared to the CG-SCF result. The SC-XLMD using Langevin thermostats slows down the latent momenta and recovers better the  auto-correlation behavior at the first several frequencies. In general, the increase of latent temperature in turn affects the charge behavior over long timescales and can be controlled \emph{via} careful application of thermostats.

\begin{figure*}[hbt!]
    \centering
    \includegraphics[scale=0.5]{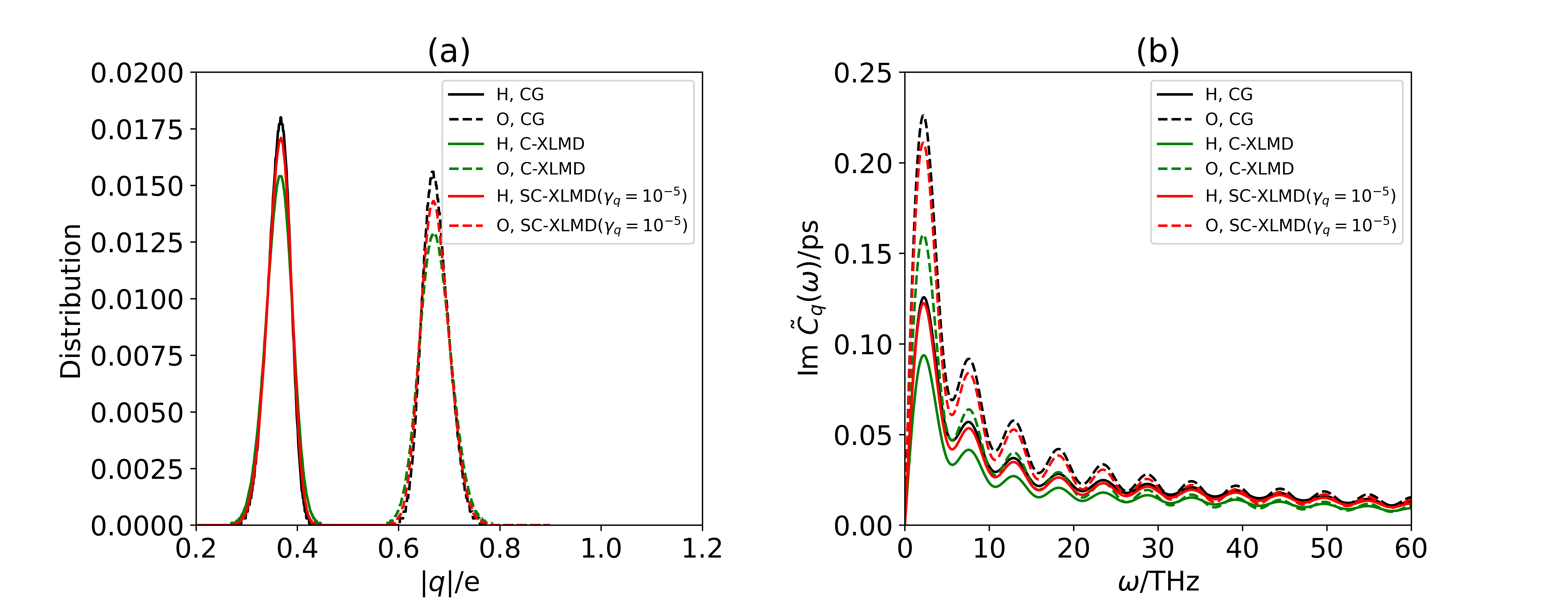}
    \caption{(a) Charge distribution function for the standard CG-SCF solution ($10^{-10}$ convergence criteria), the C-XLMD method, and the SC-XLMD method with $\gq=10^{-5}$. (b) The corresponding charge autocorrelation function in frequency domain for the three methods.}
\end{figure*}

As a comparison, previously in the iEL/0-SCF method for classical polarization, the unthermostatted latent variable are able to reproduce both the dipole distribution and dipole autocorrelation function albeit a similar increase of latent temperature is observed\cite{Albaugh17}. We note that this difference comes from the fact that the characteristic decorrelation time is more than one order of magnitude faster for CEM ($\sim10\tim$) than that for classical polarization ($\sim250\tim$). Thus the approximation used to update the real charges on the fly in the C-XLMD is intrinsically more prone to error.

\begin{figure*}[hbt!]
    \centering
    \includegraphics[scale=0.65]{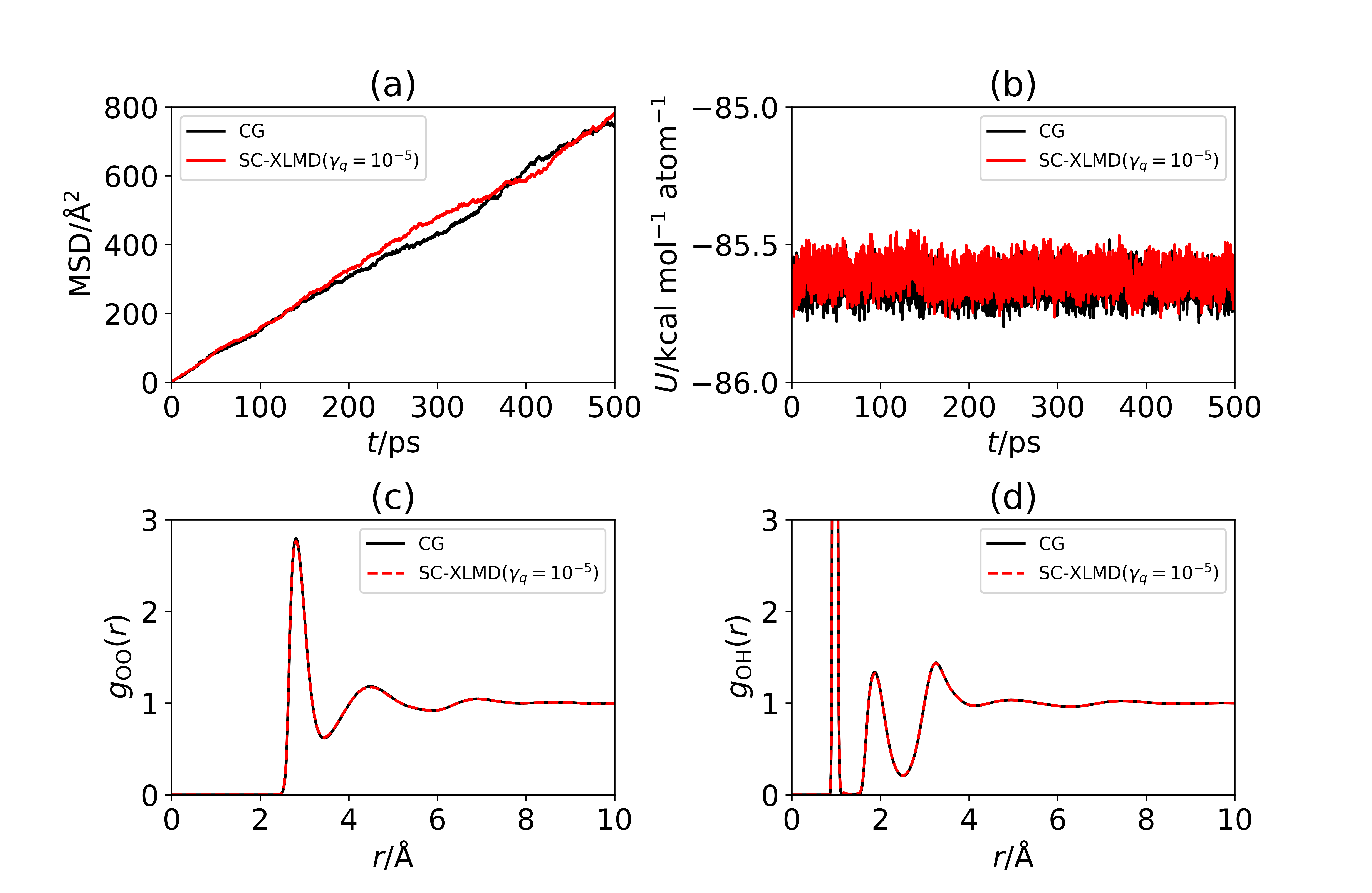}
    \caption{Comparison of methods for real charge properties of the CEM solution for bulk water. (a) Diffusion coefficient obtained by SC-XLMD ($2.52\times10^{-9}\mathrm{~m^2~s^{-1}}$) is in accord with the CG-SCF solution ($2.51\times10^{-9}\mathrm{~m^2~s^{-1}}$). (b) Potential energy obtained by SC-XLMD ($U=-85.61\pm0.04\enea$) is in accord with the CG-SCF solution ($U=-85.64\pm0.04\enea$). For the NVT simulations, we apply both a Nosé-Hoover thermostat to atomic variables and a Langevin thermostat to the latent variables. (c) radial distribution function between oxygen atoms. (d) Radial distribution function between oxygen atoms and hydrogen atoms.}
\end{figure*}
 
\begin{table}[!hbt]
\centering
\caption{The strong and weak scaling of the SC-XLMD method compared to standard CG-SCF is demonstrated in terms of both (a) different number of cores and (b) for different sizes of the ReaxFF bulk water system, showing the time cost in hours/ns calculated from the "Modify" component of a LAMMPS simulation time analysis. }
(a)
\begin{tabular}{cccc}
\toprule
$N_{\rm core}$ & CG($10^{-8}$) & CG($10^{-12}$) & SC-XLMD \\ \midrule
1               & 19.73         & 26.83          & 11.16      \\
2               & 11.42         & 15.88          & 6.32       \\
4               & 7.48          & 10.38          & 3.85       \\
8               & 5.19          & 7.66           & 2.47       \\
16              & 3.62          & 5.24           & 1.64       \\ \bottomrule
\end{tabular}
(b)
\begin{tabular}{cccc}
\toprule
$N_{\rm mlcs}$ & CG($10^{-8}$) & CG($10^{-12}$) & SC-XLMD \\ \midrule
233                 & 3.62          & 5.24           & 1.64       \\
466                 & 5.53          & 7.99           & 2.70       \\
932                 & 9.55          & 13.83          & 4.75       \\
1864                & 17.66         & 25.08          & 8.29       \\
3728                & 32.68         & 45.55          & 16.39      \\
7456                & 62.21         & 89.45          & 31.47      \\
14912               & 120.08        & 169.80         & 60.14      \\ \bottomrule
\end{tabular}
\end{table}

\textit{Other Properties.} The recommended method, SC-XLMD with $\gq=10^{-5}$, is tested for reproduction of both dynamic properties and statistical properties of the CG-SCF solution. In the NVE ensemble, the mean squared displacement (MSD) as a function of time, and subsequently the diffusion constant calculated from the MSD, is in good agreement with CG-SCF method (Figure 3a). In the NVT ensemble, the potential energy (Figure 3b) and the radial distribution function between oxygen atoms (Figure 3c) as well as between oxygen atom and hydrogen atom (Figure 3d) from SC-XLMD show excellent agreement with CG-SCF.

\textit{Benchmarks.} Finally, we show that the iteration-free extended dynamics offers significant computational cost advantages over the standard CG-SCF method. We have previously shown that $\sim70$ iterations are required to reach convergence at each timestep for the water system illustrated here, and the iEL/SCF procedure we developed for CEM was able to reduce this to $\sim20$ SCF cycles. The SC-XLMD method, by eliminating SCF cycles altogether, is found to perform better for both strong and weak scaling, i.e.: (1) the scaling with increasing number of cores (Table 1a) since there are less communication between processors to exchange the information of charges; (2) the scaling with increasing size (Table 1b) of the simulation systems since the $O(n^2)$ matrix-vector multiplication is significatly reduced. Due to these features, our implementation of the SC-XLMD approach should benefit even more with the recent optimized software implementations of LAMMPS on many-core hardware architectures.\cite {Aktulga12,Aktulga19}  

\section{Conclusion}
The extended Lagrangian approach that eliminates the self-consistent field step for polarization, iEL/0-SCF, has been extended to charge equilibration models that require the solution of two sets of linear equations for the charges under the constraint of charge conservation. By creating two latent variables of charges and chemical potential under stochastic thermodynamic control, and solving the XLMD with a holonomic constraint that preserves charge conservation, the resulting SC-XLMD is stable and maintains desired accuracy, and yields significant computational speed-ups relative to a standard  SCF solver implemented in the reference program LAMMPS. With no SCF cycles to consider, the solution for the many-body CEM forces is now commensurate with the cost of two-body fixed charge calculations, opening up ReaxFF calculations to much larger systems and longer timescales than previously possible. 

The successful formulation and application of SC-XLMD also suggests that SCF-less solutions are widely applicable to more many-body models. We have asserted previously that the iEL/0-SCF method yields satisfactory result because the characteristic decorrelation time $\tau$ for polarizable force field model is $\sim 500$ times larger than the time step, thus the iEL/0-SCF method is effectively doing SCF iterations on-the-fly. However for CEM this ratio is reduced by an order of magnitude, and yet an SCF-less solution proved viable for CEM using the SC-XLMD method. We thus look forward to reporting more SCF-less solutions for many-body potentials such as \emph{ab initio} molecular dynamics, where the characteristic decorrelation time is similar to the CEM (i.e., $\sim10-30\tim$).

\begin{acknowledgement}
S. T. thanks the University of California Education Aboard Program (UCEAP) for financial and visa support. This work was supported by the U.S. Department of Energy, Office of Science, Office of Advanced Scientific Computing Research, Scientific Discovery through Advanced Computing (SciDAC) program (S.T., I.L., L.L., T.H-G.), and by the National Science Foundation under grant DMS-1652330 (D.A. and L.L.). The authors thank NERSC for computational resources. The authors thank S. Y. Cheng for useful discussions.
\end{acknowledgement}
\bibliography{ref}
\end{document}